\documentclass[review]{elsarticle}
\usepackage{amsmath,amssymb}
\usepackage{graphicx}
\usepackage{hyperref}

\journal{Journal of \LaTeX\ Templates}

\begin{document}

\begin{frontmatter}

\title{Asymmetric behavior of the B$(E2 \uparrow; 0^+ \rightarrow 2^+)$ values in $^{104-130}$Sn and Generalized Seniority}

\author{Bhoomika Maheshwari\corref{mycorrespondingauthor}}

%
\cortext[mycorrespondingauthor]{Corresponding author}
\ead{bhoomika.physics@gmail.com}
\author{Ashok Kumar Jain}
\address{Department of Physics, Indian Institute of Technology, Roorkee-247667, India.}
\author{and Balraj Singh}
\address{Department of Physics and Astronomy, McMaster University, Hamilton, Ontario-L8S 4M1, Canada.}

\begin{abstract}

We present freshly evaluated B$(E2\uparrow;0^+\rightarrow2^+)$ values across the even-even Sn-isotopes which confirm the presence of an asymmetric behavior as well as a dip in the middle of the full valence space. We explain these features by using the concept of generalized seniority. The dip in the B$(E2)$ values near $^{116}$Sn is understood in terms of a change in the dominant orbits before and after the mid shell, which also explains the presence of asymmetric peaks in the B$(E2)$ values. This approach helps in deciding the most active valence spaces for a given set of isotopes, and single out the most useful truncation scheme for Large Scale Shell Model (LSSM) calculations. The LSSM calculations so guided by generalized seniority are also able to reproduce the experimental data on B$(E2)\uparrow$ values quite well. 

\end{abstract}

\begin{keyword}
\texttt{$2_1^+$ states, Sn-isotopes, Generalized seniority, Even-tensor $E2$ transitions}
\end{keyword}

\end{frontmatter}


\section{Introduction}

The Sn-isotopes present the longest available isotopic chain between two doubly magic nuclei from $^{100}$Sn to $^{132}$Sn, and beyond. They provide a very useful data set to explore nuclear structural properties across the $N=50$ to $82$ shell, and also to test the realistic effective interactions ~\cite{simpson,maheshwari}. The concept of seniority~\cite{racah} has been successfully used in the past as an important tool to understand the behavior of the semi-magic nuclei, from the particle number independent energies to the parabolic B$(E2)$ variation~\cite{casten,maheshwari1}. We have recently presented a simple microscopic scheme for the generalized seniority in multi-j degenerate orbits, where we find that the electric transitions, for both the even and the odd tensors, behave similar to each other. This enabled us to find for the first time a new kind of seniority isomers due to the odd-tensor $E1$ transitions in the Sn-isotopes~\cite{maheshwari1}, and gave us chance to explain the behavior of $E3$ transitions in Sn-isotopes as well~\cite{maheshwari2}. We use the same approach in this paper to understand the behavior of the B$(E2 \uparrow; 0^+ \rightarrow 2^+)$ values in the Sn-isotopes.

The nearly constant energy of the first $2^+$ states from $^{100}$Sn to $^{130}$Sn suggests that seniority may be a good quantum number in these isotopes. In a theoretical study, Sandulescu $et$ $al.$ ~\cite{sandulescu} compared the energies of the yrast generalized seniority states with shell model states for $^{104-112}$Sn, and concluded that a model space with seniority greater than 2 is probably necessary. Several groups, both theoretical and experimental, have been studying the B$(E2 \uparrow; 0^+ \rightarrow 2^+)$ variation in the Sn-isotopes. It was expected that the B$(E2)$ values would show a parabolic variation with a peak in the middle as predicted by the seniority scheme applied to the full valence space. Earlier measurements also supported the same. However, Jungclaus $et$ $al.$~\cite{jungclaus} have recently reported new measurements which deviate from the expected parabolic behavior in the middle at $^{116}$Sn, where they find a dip rather than the expected peak~\cite{ekstrom}, and attributed it to a reduced collectivity in the middle. Morales $et$ $al.$~\cite{morales}, however, explained this minimum in terms of the different rates of filling of the orbits by using a generalized seniority approach~\cite{arima,talmi,shlomo}. 

More recently, Doornenbal $et$ $al.$~\cite{doornenbal} have reported a new measurement of B$(E2)$ in $^{104}$Sn; they have also compared the results from the large scale shell model (LSSM) and quasiparticle random phase approximation (QRPA), and other theoretical calculations~\cite{morales,ansari,ansari1,back,bader} with the available experimental data in the Sn-isotopes~\cite{jungclaus,ekstrom, radford,allmond,banu,cederkall,vaman,doornenbal1,kumar,guastalla,bader}.

Iudice $et$ $al.$~\cite{iudice} have tried to reproduce the asymmetry of B$(E2)$ parabolas by using the quasi-particle phonon model (QPM), and explore its origin in the evolution of single particle energies and polarization of the $N=Z=50$ core. On the other hand, Jiang $et$ $al.$~\cite{jiang} also presented a theoretical framework of nucleon pair approximation (NPA), where they used two sets of single particle energies, two-body interaction parameters and effective charges, one each for A $\le 116$ and A $\ge 116$, in order to reproduce the B$(E2)$ data with two asymmetric parabolas and a minimum at $^{116}$Sn. In fact, some doubts have also been raised about the existence of the minimum in the experimental B$(E2)$s as several measurements with significant variations exist.
 
The purpose of this paper is to perform a systematic study of the B$(E2)$ values in even-even Sn-isotopes for the first excited $2^+$ state in the framework of generalized seniority in multi-j environment. We found it necessary to re-evaluate the measured B$(E2)$ data for Sn-isotopes, in view of the large number of measurements differing from each other, even though a 2015 update of the B$(E2)$ values has just been published~\cite{pritychenko}. We then use a simple microscopic approach based on quasi-spin scheme~\cite{kerman,helmers} applied to the generalized seniority in many-j degenerate orbits, and calculate the reduced transition probabilities in the Sn-isotopes by a simple formula, which has been presented in our recent paper~\cite{maheshwari1}. These generalized seniority calculations, which are very handy in nature, reproduce the overall experimental trend with a minimum in the middle, and provide direct information about the orbits involved before and after the mid-shell. The seniority-guided shell model calculations are then used to support these results, which also reproduce the minimum at the middle and explain the asymmetric peaks in the B$(E2)$ values quite well. Our generalized seniority and seniority guided LSSM calculations, hence, support the understanding of two asymmetric parabolas in terms of filling of different orbits, before and after the middle, and support the results of Morales $et$ $al.$~\cite{morales}.
  
The paper is organized in four sections. We present the evaluated B$(E2)$ values for Sn-isotopes in section II. Section III presents a brief theoretical framework about the generalized seniority scheme, including the formulas used in the calculations. Section IV gives all the details of the calculations and results for both generalized seniority and seniority guided LSSM calculations. Section V summarizes the present work.  

\section{Evaluated B(E2) data}
Different experimental techniques often lead to the different B$(E2)$ values in $^{104-130}$Sn isotopes. A recently published table of B$(E2)$ values presents the evaluated data~\cite{pritychenko}, available from all the sources, and lists the recommended values. Allmond $et$ $al.$~\cite{allmond1} recently published their B$(E2)$ measurements in Sn-isotopes, which were not available at the time of this update. Besides this, the table~\cite{pritychenko} also missed a paper~\cite{orce} having measurements for $^{116,118}$Sn isotopes. Other experimental works are already included and listed in~\cite{pritychenko}. We have, therefore, reevaluated the new/missing data to incorporate these changes and presented the evaluated B$(E2 \uparrow; 0^+ \rightarrow 2^+)$ values in Table I, based on the data available until now. Most of the values are found to be almost consistent with the recent update of B$(E2)$ values~\cite{pritychenko} with slight modifications, except for $^{128,130}$Sn which remain unchanged. 


\begin{table}[htb]
\caption{\label{tab:table1}Evaluated reduced transition probabilities B$(E2 \uparrow; 0^+ \rightarrow 2^+)$, in units of $e^2b^2$, for $^{104-130}$Sn isotopes. The star $\star$ signifies that the evaluated value is same as in the recently published table of B$(E2)$s~\cite{pritychenko}. The papers listed in~\cite{pritychenko} along with the recent papers of Allmond $et$ $al.$~\cite{allmond1} and Orce $et$ $al.$~\cite{orce} have been used in the evaluation.}

\centering
\vspace{0.3cm}
\begin{tabular}{c c}
\hline\hline
Isotope   &  B$(E2)\uparrow$ $e^2b^2$\\
\hline
\vspace{0.3cm}\\
$^{104}$Sn &0.176(26) \\

$^{106}$Sn &0.209(39) \\

$^{108}$Sn &0.224(19) \\

$^{110}$Sn &0.226(22) \\

$^{112}$Sn &0.2362(70) \\

$^{114}$Sn &0.2212(73) \\

$^{116}$Sn &0.2030(45) \\

$^{118}$Sn &0.2055(34) \\

$^{120}$Sn &0.2002(37) \\

$^{122}$Sn &0.1894(55) \\

$^{124}$Sn &0.1631(33) \\

$^{126}$Sn &0.1273(72) \\

$^{128}$Sn &0.0771(38)$\star$ \\

$^{130}$Sn &0.023(5)$\star$ \\
\hline
\hline
\end{tabular}
\centering
\end{table}
In this evaluation, we have not used the experimental values from the $(e,e')$ method and any other model-dependent method. We have also used a minimum uncertainty of 4$\%$ in a certain experiment. All the values are weighted averaged values. Note that the values from Allmond $et$ $al.$~\cite{allmond1} and Jungclaus $et$ $al.$~\cite{jungclaus}, are systematically different from each other, particularly in the middle, though both the groups show a dip in the middle and support the presence of two asymmetric parabolas. So far, no B$(E2)$ measurement exists for $^{102}$Sn.

\section{Theoretical Framework}

We use the generalized seniority scheme for multi-j degenerate orbits to calculate the B$(E2)$ values in the Sn-isotopes, already presented in our recent paper~\cite{maheshwari1}. The B$(E2)$ transition probabilities between the initial $J_i$ and final $J_f$ states in multi-j case can be obtained, by defining $\tilde{j}=j \otimes j' ....$ with the total pair degeneracy $\Omega= \frac{1}{2}(2 \tilde{j} +1)= \frac{1}{2} \sum \limits_j (2j+1)$, by using the formula,
\begin{eqnarray}
B(E2) \uparrow=\frac{5}{2J_i+1}|\langle \tilde{j}^n v l J_f || \sum_i r_i^2 Y^{2}(\theta_i,\phi_i) || \tilde{j}^n v' l' J_i \rangle |^2
\end{eqnarray}
where the reduced matrix elements in the $\tilde{j}^n$ configuration can be reduced to the  configuration for seniority changing $\Delta v = 2$ transitions from the equation~\cite{maheshwari1}, 
\begin{eqnarray}
\langle \tilde{j}^n v l J_f ||\sum_i r_i^2 Y^{2}|| \tilde{j}^n v\pm 2 l' J_i \rangle  = \Bigg[ \sqrt{\frac{(n-v+2)(2\Omega+2-n-v)}{4(\Omega+1-v)}} \Bigg] \nonumber\\ \langle \tilde{j}^v v l J_f ||\sum_i r_i^2 Y^{2}|| \tilde{j}^v v\pm 2 l' J_i \rangle 
\end{eqnarray}

Note that the coefficients in the square brackets in the multi-j case, are similar to the well known single-j case, due to the simple incorporation of the multi-j by defining $\tilde{j}=j \otimes j' ....$.  The coefficient decides the behavior of the B$(E2)$ values, which depends only on the particle number $n$, the generalized seniority $v$ and the corresponding total pair degeneracy $\Omega$ in the multi-j environment. 

This means that the B$(E2)$ transition probabilities will show a parabolic behavior with a maximum at the mid-shell for $\Delta v = 2$, seniority changing transitions as is the situation in $(0^+ \rightarrow 2^+)$ transitions. We use the generalized seniority formula in the following to calculate the B$(E2 \uparrow)$s. We also obtain the B$(E2 \uparrow)$s by using the LSSM calculations, using the valence spaces guided by the generalized seniority scheme. These LSSM calculations support the generalized seniority results very well, with a minimum in the middle and two asymmetric parabolas. We present details of the calculations in the next section. 

\section{Calculations and discussion}

\begin{figure}
\includegraphics[width=13cm,height=11cm]{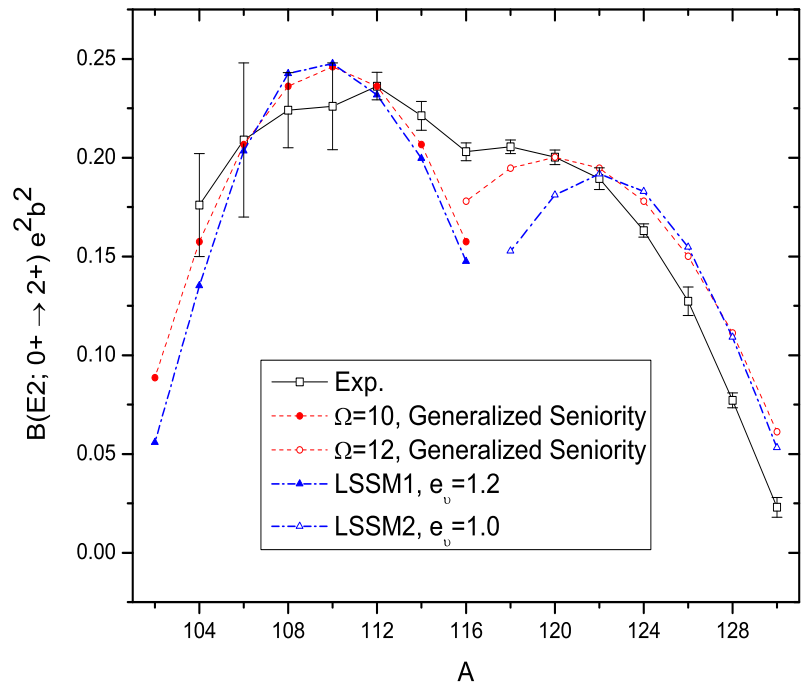} 
\caption{\label{fig:be2}(Color online) Comparison of the experimental B$(E2 \uparrow; 0^+ \rightarrow 2^+)$ values and the generalized seniority and seniority guided large scale shell model (LSSM) calculations in even-even Sn-isotopes. Evaluated data are based on the experimental measurements ~\cite{pritychenko,allmond1,orce}.} 
\end{figure}

We plot our calculated results in Fig.~\ref{fig:be2}, using dashed and dash-dotted lines for generalized seniority and seniority guided LSSM calculations, respectively. For the generalized seniority calculations, we divide available valence space of g$_{7/2}$, d$_{5/2}$, h$_{11/2}$, d$_{3/2}$ and s$_{1/2}$ orbits in the Sn-isotopes into two parts: (a) $\tilde{j}= g_{7/2} \otimes d_{5/2} \otimes d_{3/2} \otimes s_{1/2}$, $\Omega=10$ and (b) $\tilde{j}= d_{5/2} \otimes d_{3/2} \otimes s_{1/2} \otimes h_{11/2}$, $\Omega=12$. We take $^{100}$Sn as a core for $\Omega=10$, a natural choice, while $^{108}$Sn as a core for $\Omega=12$ due to the g$_{7/2}$ orbit which is now full. For dividing the valence space, we use the fact that the h$_{11/2}$ orbit mainly dominates after the mid-shell $(^{116}Sn)$, while the g$_{7/2}$ orbit completely freezes on reaching $^{116}$Sn. We then calculate the B$(E2)$ values by using the generalized seniority scheme, and obtain two asymmetric parabolas as shown in Fig.~\ref{fig:be2}. 

It is interesting to note that the two parabolas would cross over each other at $^{116}$Sn (if extrapolated in Fig.~\ref{fig:be2}), and naturally support a minimum at the middle. Therefore, the minimum of B$(E2)$ values at $^{116}$Sn surely does not correspond to any reduced collectivity as suggested earlier by Jungclaus $et$ $al.$~\cite{jungclaus}. It only describes the change in the filling of the orbits before and after the middle, and also corresponds to the location, where g$_{7/2}$ freezes out, and h$_{11/2}$ takes over. These calculations, hence, explain the overall experimental trend quite well, and also provide a direct cue for the configuration involved and their influence on the B$(E2)$ values. It may also be concluded that proper configuration mixing is required to understand the origin of the excited states in the semi-magic nuclei; the first excited $2^+$ states of the Sn-isotopes in the present case.

We further use this generalized seniority scheme to single out the active valence spaces and resultant truncations for LSSM calculations. We divide the LSSM calculations also in two parts: first we take $^{100}$Sn as a core with the multi-j configuration $\tilde{j}= g_{7/2} \otimes d_{5/2} \otimes d_{3/2} \otimes s_{1/2}$ (LSSM1) and second we choose $^{108}$Sn as a core with the multi-j configuration $\tilde{j}= d_{5/2} \otimes d_{3/2} \otimes s_{1/2} \otimes h_{11/2}$(LSSM2). The neutron single particle energies have been taken as $-10.6089$, $-10.2893$, $-8.7167$, $-8.6944$, $-8.8152$ MeV for the available 0g$_{7/2}$, 1d$_{5/2}$, 1d$_{3/2}$, 2s$_{1/2}$ and 0h$_{11/2}$ valence orbits. The harmonic oscillator potential was chosen with an oscillator parameter of $\hbar \omega =45A^{-1/3}-25A^{-2/3}$. Only two sets of the neutron effective charge have been used, $e_\nu=1.2$ in first valence space and $e_\nu=1$ in the second valence space. We have used the Nushell code~\cite{brown} for the large scale shell model calculations along with the SN100PN~\cite{brown1} effective interaction. The seniority guided shell model calculations reproduce the experimental data quite well, as shown in Fig.~\ref{fig:be2}. This also suggests that the breaking of $Z=N=50$ core is not so crucial to understand the origin of first excited $2^+$ states in Sn-isotopes. These calculations strongly validate the interpretation provided by generalized seniority about the truncations and configuration mixings needed in the generation of these $2_1^+$ states. The generalized seniority scheme applied to B$(E2 \uparrow; 0^+ \rightarrow 2^+)$ values thus provides a simple way to handle the LSSM calculations, particularly where the dimensions become very large.
 
\begin{figure}
\includegraphics[width=13cm,height=11cm]{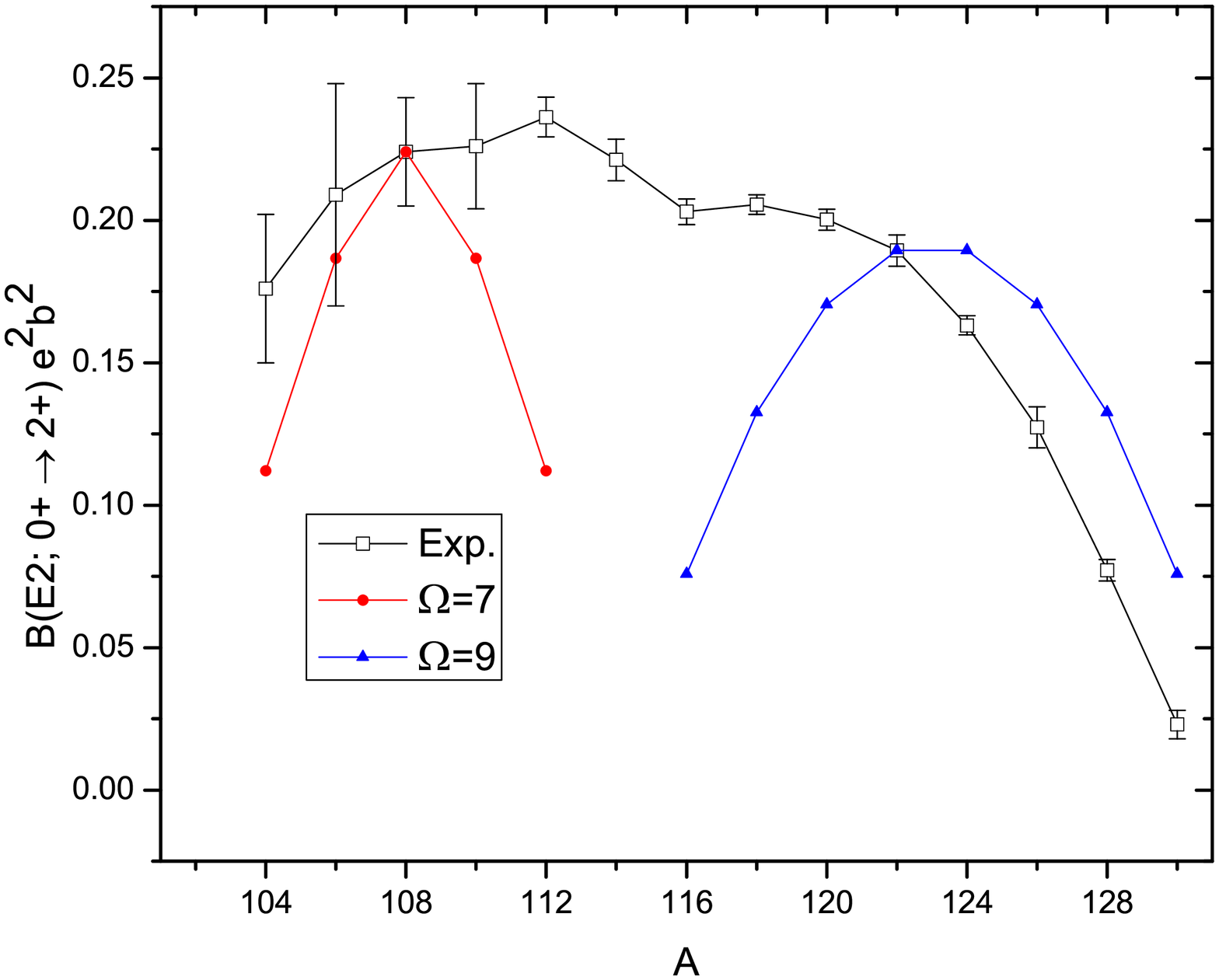}
\caption{\label{fig:be21}(Color online) Same as Fig.~\ref{fig:be2}, but the generalized seniority calculations using $\Omega = 7$, and $9$ values, corresponding to $g_{7/2} \otimes d_{5/2}$ and $h_{11/2} \otimes d_{3/2} \otimes s_{1/2}$ valence spaces, respectively, in Sn-isotopes. } 
\end{figure}

For comparison, we have also calculated the B$(E2)$ values by using $\Omega=7$ with $^{100}$Sn core and $\Omega=9$ with $^{114}$Sn core in the generalized seniority and plotted these results in Fig.~\ref{fig:be21}, along with the evaluated experimental trend. Here $\Omega=7$ and $9$ correspond to $\tilde{j}= g_{7/2} \otimes d_{5/2}$ and $\tilde{j}= d_{3/2} \otimes s_{1/2} \otimes h_{11/2}$ mixed configurations, respectively. The calculated results are quite far from the experimental data, and exhibit an obvious gap at the $^{114}$Sn isotope. One can, therefore, infer that the mixing of other orbits (as included in $\Omega=10$ and $12$) is essential for a complete understanding of these B$(E2)$ values and the asymmetric parabolas. Though the contribution of the d$_{3/2}$ and s$_{1/2}$ remains less than that for the $g_{7/2}$ and $d_{5/2}$ orbits before $^{116}$Sn, it does affect the values of the reduced transition probabilities and also the corresponding wave functions of the first excited $2^+$ states, while $h_{11/2}$ can be ignored. On the other hand, the major contribution of the $h_{11/2}$ starts after the mid-shell along with the $d_{5/2}$, $d_{3/2}$ and $s_{1/2}$ orbits, while $g_{7/2}$ can be frozen out. Though the isotopes around $^{116}$Sn may need a mixing of both the $g_{7/2}$ and $h_{11/2}$ orbits in the resultant wave functions, this results in a very large dimension in the LSSM calculations and could not be carried out by us.

Hence, the seniority generalized for multi-j orbits, provides a direct understanding of the configuration mixing, and the truncations in the Sn-isotopes. The seniority guided shell model calculations confirm the same. We, therefore, conclude that the dip in the middle, neither corresponds to the reduced collectivity, nor supports any core-excited structure in any significant way. The dip is only due to the different sets of active orbits and their different rate of mixings, which finally leads to the asymmetric B$(E2)$ parabolas. These results also agree with the conclusions of Morales $et$ $al.$~\cite{morales}.   

\section{Conclusion}

In conclusion, we have re-evaluated the B$(E2 \uparrow; 0^+ \rightarrow 2^+)$ transition probabilities in the Sn-isotopes and provided a set of most acceptable values. The evaluated data confirm the existence of a dip in the middle of full valence space. We, then, use them in the generalized seniority scheme for multi-j degenerate orbits. These simple calculations reproduce the experimental data quite well along with the dip in the middle. This also works as a guide to fix the orbits involved in configuration mixing. This information turns out to be very useful in deciding the truncations in the LSSM calculations, which again reproduce the experimental trend very well. The dip in the middle, therefore, corresponds to a competition between the orbits in $N=50-82$ valence space, and pinpoints the location where the role of the $g_{7/2}$ orbit is taken over by the $h_{11/2}$ orbit. It is remarkable that the calculated values from our formulation match with the LSSM results very well. This confirms the effectiveness of our formulation based on the generalized seniority. We also conclude that the minimum in the middle does not correspond to a reduced collectivity as also pointed out by Morales $et$ $al.$~\cite{morales}. 

\section*{Acknowledgments}

Financial support from the Ministry of Human Resource Development (Government of India) is gratefully acknowledged.

\end{document}